\documentclass[titlepage,12pt]{utarticle}  
\usepackage{amssymb}
\usepackage{floatflt,graphicx} %useful for figures
\usepackage{cite}
\renewcommand{\baselinestretch}{1.3} \normalsize

\numberwithin{equation}{section}

\newdimen\tableauside\tableauside=1.0ex
\newdimen\tableaurule\tableaurule=0.4pt
\newdimen\tableaustep
\def\phantomhrule#1{\hbox{\vbox to0pt{\hrule height%
	\tableaurule width#1\vss}}}
\def\phantomvrule#1{\vbox{\hbox to0pt{\vrule width%
	\tableaurule height#1\hss}}}
\def\sqr{\vbox{%
  \phantomhrule\tableaustep
\hbox{\phantomvrule\tableaustep\kern\tableaustep\phantomvrule\tableaustep}%

\hbox{\vbox{\phantomhrule\tableauside}\kern-\tableaurule}}}
\def\squares#1{\hbox{\count0=#1\noindent\loop\sqr
  \advance\count0 by-1 \ifnum\count0>0\repeat}}
\def\tableau#1{\vcenter{\offinterlineskip
  \tableaustep=\tableauside\advance\tableaustep by-\tableaurule
  \kern\normallineskip\hbox
    {\kern\normallineskip\vbox
      {\gettableau#1 0 }%
     \kern\normallineskip\kern\tableaurule}%
  \kern\normallineskip\kern\tableaurule}}
\def\gettableau#1 {\ifnum#1=0\let\next=\null\else
  \squares{#1}\let\next=\gettableau\fi\next}

\def\cN{{\cal N}}

\newfont{\goth}{eufm10 scaled \magstep1}

\def\a{\alpha}

\def\c{\gamma}
\def\d{\delta}

\def\vf{\varphi}

\def\th{\theta}

\def\del{\partial}

\def \ys {{y\kern-.5em / \kern.3em}}

\def\bd{\begin{document}}
\def\ed{\end{document}}
\def\ba{\begin{array}}
\def\ea{\end{array}}
\def\bea{\begin{eqnarray}}
\def\eea{\end{eqnarray}}
\def\fft#1#2{{#1 \over #2}}
\def\be{\begin{equation}}
\def\ee{\end{equation}}
\newcommand{\eq}[1]{(\ref{#1})}
\def\eqs#1#2{(\ref{#1}-\ref{#2})}
\def\det{{\rm det\,}}
\def\tr{{\rm tr}}
\newcommand{\ho}[1]{$\, ^{#1}$}
\newcommand{\hoch}[1]{$\, ^{#1}$}
\def\ra{\rightarrow}
\def\uha{{\hat {\underline{\a}} }}
\def\uhc{{\hat {\underline{\c}} }}

\def \Om {\Omega}
\def \bfd {{\bf d}}
\def \del {\partial}
\def \eps {\epsilon}
\def \Z {{\bf Z}}
\def \xb {\bar{x}}
\def \lg {\langle}
\def \rg {\rangle}
\def \Omt {\tilde \Omega} 
\def\l{\left}
\def\r{\right}
\def\kt{\widetilde{k}}
\def\Kt{\widetilde{K}}
\def\co{\rm cos}
\def\si{\rm sin}
\def\vft{\widetilde{\varphi}}

\begin{document}

\preprint{
UTTG--12-00\\
{\tt hep-th/0007082}\\
}
\title{
Non-Supersymmetric $SO(3)$-Invariant Deformations
of $\cN=1^\star$ Vacua and their Dual String 
Theory Description
}
\author{\hspace{3mm} Frederic Zamora
 \oneaddress{
    Theory Group, Physics Department\\
    University of Texas at Austin\\
    Austin TX 78712 USA.\\
{~}\\
\email{zamora@zerbina.ph.utexas.edu}}
}
\date{July 11, 2000}

\Abstract{
We study the $SO(3)$-invariant relevant deformations
of $\cN=4$ $SU(N)$ gauge theory using the methods 
of Polchinski and Strassler. 
We present the region of parameter space where 
the non-supersymmetric vacuum is still described by  
stable ``dielectric'' five branes within the 
supergravity approximation.
}

\maketitle
%%%%%%%%%%%%%%%%%%%%%%%%%%%%%%%%%%%%%%%%%%%%%%%%%%%%%%%%%%%%%%%%%%%

\section{Introduction}\label{sec:intro}

Via the AdS/CFT correspondence
\footnote{See for instance \cite{Aharony:1999ti} 
for a review.}
one can study relevant deformations of four 
dimensional $\cN=4$ $SU(N)$ gauge theory at strong 't Hooft coupling.
Using this technique we have obtained new insights
about RG flows \cite{Distler:1998gb,Girardello:1998pd,
Balasubramanian:1999jd,Freedman:1999gp,
Girardello:1999bd,Skenderis:1999mm,
Klebanov:1999rd,deBoer:1999xf,Lowe:2000ir,Oh:2000ez,
Pilch:2000ue,Brandhuber:2000ct,Pilch:2000fu,
Angelantonj:1999qg,Angelantonj:2000kh,Bianchi:2000vb}, 
IR fixed points and the c-theorem 
\cite{Alvarez:1998wr,Girardello:1998pd,Karch:1999pv,
Freedman:1999gp,Anselmi:2000fu,Pilch:2000ej}
and possible chiral symmetry breaking \cite{Distler:1999tr}.
But we have also learned that,
unless the perturbations are fine tuned 
to end up at some stable IR fixed point
(which up to now always seems to require 
some unbroken supersymmetry),
the spacetime geometry usually 
develops a naked singularity deep into the bulk
\footnote{See \cite{Gubser:2000nd} for a discussion
on these singularities.}.
Close to it, which corresponds to the 
field theory low energy physics, supergravity 
ceases to be valid and stringy corrections become important.

In the case of $\cN=1$ supersymmetric mass deformations,
the so called ${\cN=1^\star}$ field theory,
Polchinski and Strassler \cite{Polchinski:2000uf}
made a precise proposal for the string theory 
resolution of the singularity:
it is replaced by a five brane configuration
\footnote{See \cite{Aharony:2000cw,
Bena:2000zb,Bena:2000fz,Bena:2000va} for
some recent extensions of \cite{Polchinski:2000uf}.}.
In principle, for any $\cN=1^\star$ field theory vacuum,
there is a dual description in terms of a 
type IIB background involving several 
``dielectric'' five branes with D3 charges.
The brane configurations create some
type IIB string theory boundary conditions,
and we know how to read the dictionary 
between particular boundary conditions
and field theory bare Lagrangians.
To obtain $\cN=1^\star$ Lagrangians, we need
non-trivial boundary conditions for the 
magnetic three-form.
Their sources are five branes with non-zero electric dipoles.

Once this set up was established, 
Polchinski and Strassler proceeded
to find a computable dual string theory description 
of the $\cN=1^*$ field theory vacua. 
Their approach was to take 
the most similar vacuum in the $\cN=4$ Coulomb branch,
a continuous distribution of D3 branes with the topology
of a two sphere, and compute the linearized perturbed 
background caused by the introduction 
of non-vanishing boundary 
conditions to the three-form (the fermion masses).
It results that, in certain regions of parameter space,
the linearized solution is enough 
to have a valid description.
In presence of the magnetic three-form, the D3 branes
are replaced by several five branes via the Myers effect
\cite{Myers:1999ps}.
Then, the whole problem is reduced 
to find the five brane configuration
that produces the perturbed type IIB background.
A good criteria is to look for those configurations 
which minimize the five brane effective action
in such a background. It was
realized in \cite{Polchinski:2000uf} that if the ratio of 
five and three brane charge densities is very small,
the linearized solution becomes a valid approximation
around the five brane action minima.
\footnote{For completeness, we should mention that there is a
second order term (in the fermion masses) 
which should be included in the five brane action.
But supersymmetry allows one to obtain it 
without having to compute it explicitly.}

Besides this computational condition,
which we refer as the PS approximation,
there are the usual constraints for the 
classical supergravity regime: small curvature 
in Planck and string units.
Now, we have to be 
more careful, because the brane configuration
produces non-constant curvature and dilaton backgrounds.
In fact, close to the five branes, we have to interpolate
from the three brane distribution metric to the near-shell
five brane metric with D3 charge. 
For instance, for the case of a 
D5 brane, the PS approximation corresponds to 
\begin{equation}
1<<\frac{N}{g}<<N^2
\end{equation}
and for a NS5 brane, we have the $S$-dual condition
\begin{equation}
1<<gN<<N^2 \,.
\end{equation}

As it was already noticed in \cite{Polchinski:2000uf}, 
the regime of validity of their approximation 
can be extended to certain
non-supersymmetric directions in the parameter space.
In this paper, we look for those non-supersymmetric directions 
which preserve $SO(3)$ invariance and 
analyse the region which is still within the regime 
of validity of the PS approximation. We will complete the 
determination of the effective potential started 
in \cite{Polchinski:2000uf}
and study the resulting vacuum structure.

This paper grew out of an attempt \cite{DSZ}
to apply the methods of \cite{Polchinski:2000uf}
to the non-supersymmetric theory with dynamically-broken chiral symmetry,
discussed in \cite{Distler:1999tr}
(in the notation below, this is the theory with
$m= \mu^2 =0$, and $m'\neq 0$). Unfortunately, the PS approximation proved
not to be valid for that theory. Hence the motivation for the present
work: to find the precise realm of validity of the PS approximation within
the space of non-supersymmetric, $SO(3)$-invariant extensions of 
\cite{Polchinski:2000uf}.

\newpage
%%%%%%%%%%%%%%%%%%%%%%%%%%%%%%%%%%%%%%%%%%%%%%%%%%%%%%%%%%%%%%%%%%

\section{\boldmath{$\cN=0$ $SO(3)$}-invariant Five Brane Potential}

\subsection{PS approximation}

We begin by presenting the relevant effective potential for 
a five brane probe within the PS approximation:
\begin{equation}
\begin{split}
-S =\frac{\mu_5 V}{g} \l\{ |M|^2 \int_{S^2}d^2\xi
\frac{\sqrt{{\rm det}G_\parallel}{\rm det}G_\perp}
{4\pi\alpha'\sqrt{{\rm det}F_2}}
+\frac{\sqrt 2}{3}\int_{S^2}{\rm Im}\l(
{\overline M}T_{mnp}x^mdx^n\wedge dx^p\r)
\r.
\\
\l.+\pi\alpha' \int_{S^2}F_2\l(T_{i\bar{j}\bar{k}}
\overline{T}_{\bar{l}jk}z^i\overline{z}^{\bar l} 
+\mu_{mn}x^mx^n\r)\r\} \,.
\label{5pot_2}
\end{split}
\end{equation}
In this subsection we will explain
what are the fields entering into this potential, 
where it comes from, and when it is a good approximation.

First, we introduce the relevant fields 
entering in \eqref{5pot_2}
\footnote{We follow the conventions of 
\cite{Polchinski:2000uf} faithfully.}.
Since the fermion mass matrix transform in the ${\bf 10^\star}$
of $SU(4)$, it can be represented by an imaginary anti-self-dual
three form
\begin{equation}
\star_6 T_3 =-i T_3 \,,
\end{equation}
where $\star_6$ acts in the six dimensional transverse space, 
parametrized by $x^m$ ($m=4,...9$), with 
respect to the flat Euclidean metric $\d_{mn}$.
Because $\bf 10^\star$ is a complex representation,
it is convenient to use the complex coordinates
\begin{equation}
z^i=\frac{x^{i+3}+i x^{i+6}}{\sqrt 2} \ ,\quad i=1,2,3 \,.
\end{equation}
For the $SO(3)$ invariant case, where $m$ is the $\cN=1$ 
preserving mass of the three chiral superfields, and $m'$ 
is the gaugino mass (which breaks supersymmetry softly),
the anti-self-dual form becomes
\begin{equation}
T_3 = m\epsilon_{ijk} dz^i\wedge d{\overline z}^j \wedge
 d{\overline z}^k + m' dz^1\wedge dz^2\wedge dz^3 \,.
\end{equation}
The ten dimensional (string) metric is of the form
\begin{equation}
ds^2 = \frac{1}{\sqrt Z}dx_{\parallel}^2
+\sqrt{Z} dx^m dx^m \,.
\label{10d_metric}
\end{equation}
When $r^2=x^m x^m \to\infty$, we have the 
AdS$_5\times S^5$ boundary conditions
\begin{subequations}
\begin{align}
&Z\to\frac{4\pi{\alpha'}^2}{r^4} \,,
\\
&e^\Phi \to g \ ,\quad C \to \frac{\th}{2\pi} \quad 
({\rm constant\ dilaton\ and\ axion})\,,
\\
&F_5 \to d\chi_4 +\star d\chi_4 \,, \quad 
\chi_4=\frac{1}{gZ} dx^0\wedge dx^1\wedge dx^2 \wedge dx^3 \,,
\\
&\star_6 G_3 -iG_3 \to -i\frac{2\sqrt{2}}{g}ZT_3 \,,
\label{6form}
\end{align}
\end{subequations}
where $G_3=F_3-\tau H_3$ and $\tau=C+ie^{-\Phi}$
is the usual complex field theory coupling;
$F_3$ and $H_3$ are the RR and NS-NS three-form 
field strengths, respectively.

We are considering a five brane probe,  
with D3 brane charge $n$,
in type IIB background 
solution of the supergravity equations of motion
subject to the above boundary conditions.
Due to the particular expression for the non-zero six-form
potentials determined by \eqref{6form}, 
the five brane world-volume is taken 
to be $\bf{R^4\times S^2}$, with the two sphere
in the six dimensional transverse space.
$G_\parallel$ and $G_\perp$ correspond to the 
target spacetime metrics in the directions 
$\bf{R}^4$ and ${\bf S^2}$, respectively.
The Born-Infeld $U(1)$ field, $F_2$, has a non-zero 
flux around the two sphere, giving the D3 brane charge,
\begin{equation}
\int_{S^2}F_2=2\pi n \,.
\end{equation}
Finally, $M=c\tau +d$ for a $(c,d)$ five brane 
and $V$ is the (regularized) $\bf{R}^4$ coordinate volume.
We will discuss on the constant matrix $\mu_{mn}$ later.

Next, we explain the origin of the terms appearing in
\eqref{5pot_2}.
The first term corresponds to 
an expansion of the Born-Infeld square-root 
term due to the small five brane metric $G_\perp$ corrections with 
respect to the dominant D3 charge given by $F_2$. 
Observe that the $Z$ factors cancel explicitly,
because of the particular ansatz 
for the metric \eqref{10d_metric}.
The second term comes from the 
linearized (in the fermion masses) six-form potential 
$B_6 c+C_6 d$.
The PS analysis showed that it is independent of the metric factor 
$Z$.

To determine the third term would require more work.
It comes from the second order solution, in the fermion
masses, for the supergravity fields.
It should be included, since posteriori analysis 
on the location of the minimum shows that this term 
is actually of the same order as the previous 
two terms in the action.
The traceless matrix $\mu_{mn}$ represents the 
$L=2$ harmonic in the transverse five dimensional 
compact space. It receives contributions form the 
dilaton, spacetime metric and five-form field strength.
When $\cN =1$ supersymmetry is unbroken, and we take the 
round ansatz $F_2=\frac{n}{2}\sin\th$ for the $SO(3)$
invariant case, the potential \eqref{5pot_2}
should be a perfect square in terms of the 
complex scalar fields $z^i$. This requirement determines
$\mu_{mn}=0$. In our case, we lack supersymmetry
and we should find some other way to determine $\mu_{mn}$.

Before doing that, let us finish this subsection  
with a comment on the validity of the 
potential \eqref{5pot_2}. A close analysis shows that the 
full five brane action has been expanded in the controlling 
parameter $g|M|^2/n$. The PS approximation consists
in going to the region of parameter space where it is 
small. Our goal is to find out what that region is 
for the $\cN=0$ $SO(3)$-invariant case and to analyze
the corresponding vacuum structure therein.

\subsection{\boldmath{$\cN=0^*$ $SO(3)$}-Invariant Deformations}

There are three complex parameters parametrizing 
the $SO(3)$-invariant relevant deformations.
The branching rules of $SU(4) \to SU(3)\times U(1)_R$ are
\begin{subequations}
\begin{align}
& {\bf 4}\to {\bf 3}_{-1/3} + {\bf 1}_1 \,,
\\
&{\bf 6}\to {\bf 3}_{2/3} +{\bf 3^\star}_{-2/3} \,,
\\
&{\bf 10^\star} \to {\bf 6^\star}_{2/3} 
+{\bf 3^\star}_{-2/3}+{\bf 1}_{-2}\,,
\\
&{\bf 20'} \to {\bf 8}_0 +{\bf 6}_{4/3}+{\bf 6^\star}_{-4/3} \,.
\end{align}
\end{subequations}
The Abelian factor $U(1)_R$ is the $R$-symmetry of the 
$\cN=1^\star$ field theory. 
The gluino, which is the $SU(3)$ singlet 
in the $\bf 4$, has $R$-charge one.
The complex singlet ${\bf 1}_{-2}$
in ${\bf 10^\star}$ corresponds to $m'$,
the supersymmetry breaking gluino mass.
The three $\cN=1$ chiral superfields are the ${\bf 3}_{2/3}$,
with the ${\bf 3}_{-1/3}$ from the 
$\bf 4$ being its fermion components.
The $\cN=1$ preserving mass matrix
corresponds to the ${\bf 6^\star}_{2/3}$.
When its three eigenvalues are equal to $m$, 
there is an unbroken $SO(3)$, which is the real embedded
subgroup of $SU(3)$, up to a global complex 
phase (the phase of $m$). 
The  $\bf 20'$ represents a traceless mass matrix
for the six scalars
and can also been understood as the 
$L=2$ harmonic $\mu_{mn}$ in \eqref{5pot_2}. 

As it is observed in \cite{Freedman:1999gp,Polchinski:2000uf}, 
the $L=2$ harmonic only enters
in the homogeneous part of the supergravity equations of motion.
The inhomogeneous term is determined by the 
anti-self-dual three form $T_3$, which in our case includes
the parameters $m$ and $m'$. 
In the case of the $\cN=1$ $SO(3)$ boundary conditions,
the particular solution gives $\mu_{mn}=0$. 
But there are two additional 
$SO(3)$ singlets in the $\bf 20'$,
which break supersymmetry when they are turned on.
This change in the boundary conditions can be 
effectively parametrized by a single
complex parameter $\mu^2$ (dimension mass square),
which determines the $L=2$ harmonic in \eqref{5pot_2}
to be 
\begin{equation}
\mu_{mn}x^m x^n = {\rm Re}[\mu^2 z^i z^i] \,.
\end{equation}
Observe that $U(1)_R$ selection rules would allow
a possible additional ${\rm Re}[m m' z^i z^i]$ term.
But $SO(4)$ symmetry when $\mu^2=0$ and $m=m'$ 
(which enters into the region of PS approximation,
see subsection 3.3) eliminates this possibility.
\footnote{We would like to thank M. Strassler for a 
discussion on this point.}

\subsection{The Potential and Its Critical Points}

Because of the $SO(3)$ symmetry, we take the same  
ansatz as in \cite{Polchinski:2000uf}
\begin{equation}
z^i=z e^i \,,
\end{equation}
where $e^i=e^i(\th,\phi)$ represents a unit three dimensional
real vector parametrizing the two sphere. Its coordinate radius
is given by $|z|/\sqrt{2}$.
Inserting this ansatz into the action \eqref{5pot_2}
we get the effective potential
\begin{equation}
\begin{split}
V_{(c,d)}(z)=-\frac{S}{V}=&\frac{4}{\pi g n(2\pi\alpha')^4}
\l\{|M|^2|z|^4 +\frac{(2\pi\alpha')n}{3\sqrt{2}}
{\rm Im}\l[3m\overline{M}z\overline{z}^2+m'\overline{M}z^3\r]\r.
\\
&\l.+\frac{n^2(2\pi\alpha')^2}{8}\l(|m|^2+\frac{|m'|^2}{3}\r)|z|^2
+\frac{n^2(2\pi\alpha')^2}{8}{\rm Re}[\mu^2 z^2]\r\} \,.
\label{5pot_3}
\end{split}
\end{equation}
Since we are in a perturbed AdS$\times S^5$ geometry,
the value of $z$ which minimizes the previous potential
should be interpreted as the location of the five brane 
along the radial direction of AdS, 
with the two sphere being an equator of the $S^5$. 

Let us introduce the dimensionless complex parameters
\begin{equation}
b=\frac{m'}{m} \ ,\quad\quad c=\frac{\mu^2}{m^2}
\end{equation}
and also write 
\begin{equation}
z= \frac{(2\pi\alpha') n m}{\sqrt{8} M}i x e^{i\varphi} \,.
\label{eq:z}
\end{equation}
Putting this into the five brane potential, we get
\begin{equation}
\begin{split}
V_{(c,d)} (x,\vf) = \left(\frac{m^4 n^2}{16\pi}\right)
\frac{n}{g|M|^2}x^2 \left\{x^2 -2 x\,{\rm Re}\left[e^{-i\vf} 
+e^{3i\vf}\frac{b}{3}\left(\frac{\overline M}{M}\right)^2\right]
\right.
\\
\left. +1+\frac{|b|^2}{3} 
-{\rm Re}\left[e^{2i\vf}c\frac{\overline M}{M}\right]\right\} \,.
\end{split}
\label{5pot}
\end{equation}

The critical points of the potential \eqref{5pot_3}
(besides $z=0$) are those real values of $\{x,\vf\}$
that satisfy the equation
\begin{equation}
x^2 + P x +Q =0 \,,
\label{eq:ext}
\end{equation}
with the (generically complex) coefficients
\begin{subequations}
\begin{align}
P=-\frac{1}{2}\left(e^{-i\vf}+2 e^{i\vf}
+e^{3i\vf} b\left(\frac{\overline M}{M}\right)^2\right) \,,
\label{eq:P}
\\
Q=\frac{1}{2}\left(1+\frac{|b|^2}{3} -e^{2i\vf}c\left(
\frac{\overline M}{M}\right)\right) \,.
\label{eq:Q}
\end{align}
\label{eq:PQ}
\end{subequations}

We could arrange several branes along different AdS radial directions.
It means they have different D3 charge
\footnote{The AdS location 
is proportional to the D3 charge, see \eqref{eq:z}.}.
From the field theory
point of view it corresponds to fuzzy (non-commutative) scalars
in a reducible representation of $SU(2)$, 
with different dimensions for several irreducible blocks.
These configurations always leave some unbroken Abelian factors
of the gauge group. They correspond to the 
Coulomb vacua. 

Consider several five branes in the bulk,
with D3 charge $n_I$ for the $(c_I,d_I)$ five brane.
The effective potential is just the sum of all of them
\begin{equation}
V_{\rm total}=\sum_I V_{(c_I,d_I)}(x_I,\vf_I) \,.
\label{V_total}
\end{equation}

Since the metric factors $Z$ cancel in the PS approximation,
we can take \eqref{V_total}
as the actual full potential in the background created
by the five branes.
Due to the breaking of supersymmetry, 
we expect a unique vacuum for 
any given point in the parameter space
({\it i.e.}, a unique five brane configuration 
is going to be selected).
Furthermore, a numerical analysis shows that 
generically the vacuum energy is negative.
As long as we confine ourselves 
to the classical PS approximation,
the criteria is to look for the brane configuration
with the most negative value of \eqref{V_total}. 

Observe that the potential \eqref{5pot} 
is proportional to the inverse of the 
small controlling parameter $g|M|^2/n$.
This is good, because the selected brane configuration
will correspond to the one where the PS approximation 
is the most optimal for each five brane involved.
In fact, due to the constraint $N=\sum_I n_I$,
the most favorable
situations are the ones containing only one five brane.
Hence, as long as we move in the region of 
parameter space corresponding to the PS approximation, 
the massive vacua are selected.

\newpage
%%%%%%%%%%%%%%%%%%%%%%%%%%%%%%%%%%%%%%%%%%%%%%%%%%%%%%%%%%%%%%%%

\section{Analysis of the \boldmath{${\cN}=0^*$} Massive Vacua}

\subsection{\boldmath{$\cN =1^*$} Massive Vacua 
and Their Dual Description}

Since the work of 't Hooft 
\cite{'tHooft:1978hy,'tHooft:1979uj,'tHooft:1981ht},
we know which are the 
possible massive phases of four dimensional 
$SU(N)$ gauge theories in which all fields transform trivially
under the center of $SU(N)$.
They are classified by index N subgroups $P$, of the lattice
$\Z_N\times\Z_N$, the group of external 
magnetic and electric charges by which this
theory might be probed
\footnote{Charges are normalized such that
$(0,1)$ represents the smallest electric charge.}. 

Later on Donagi and Witten \cite{Donagi:1996cf} 
found the integrable 
system behind the four dimensional $\cN =2$ $SU(N)$
gauge theory with a massive hypermultiplet in the adjoint
of the gauge group. They described the Coulomb branch 
moduli space of the theory 
in terms of $N$-fold covers of a genus $N$
Riemann surface $C$, over a Torus with modulus $\tau$,
the microscopic complexified gauge coupling of the underlying 
field theory. At generic points in the Coulomb branch,
the gauge group is broken to $U(1)^{N-1}$. But at some 
particular points, $r$ one-cycles of $C$ degenerate,
giving rise to $r$ massless mutually local hypermultiplets.
When the theory is perturbed to $\cN =1$ by a mass term
for the $\cN=1$ chiral superfield in the $\cN =2$ vector 
multiplet, the Coulomb branch flat directions are lifted
and only the isolated points where the Riemann surface
degenerates remain. These are the $\cN =1^*$ vacua, where the 
$\cN =1$ mass perturbation has produced the condensation 
of the $r$ hypermultiplets associated to the singular 
one-cycles of $C$. Only at the points of maximal degeneration,
where $r=N-1$, do we have a mass gap. At these points,
$C$ reduces to genus one, and the Donagi-Witten curve 
parametrizes an unramified $N$-fold cover of $C$ over 
the Torus with modulus $\tau$. 
These unramified $N$-fold covers are in one to one 
correspondence with index N subgroups $P$  of the additive 
lattice $\Z_N\times\Z_N$.

If $N$ is prime, there are only 
$1+N$ subgroups P. The one generated by $(0,1)$
represents the Higgs vacuum, where 
the smallest electrically charged particle condenses.
In terms of the dual string theory description, 
it involves one $(0,1)$ five brane with D3 charge $N$.
The remaining $N$ subgroups are generated by $(1,s)$,
$s=0,...,N-1$, corresponding to confinement or 
oblique confinement. Each of these vacua has a 
dual description in terms of a single $(1,s)$ five brane.

If $N$ is not prime, for every positive divisor $p>1$ of $N$
and every $r$ with $0\leq r \leq p-1$, there is an index N
subgroup of $\Z_N\times\Z_N$ generated by $(0,p)$ and 
$(N/p,r)$. 
In terms of the dual type IIB background, these massive 
vacua are represented either by $p$ D5 branes, each with 
$N/p$ D3 charge, or by $N/p$ NS5 branes, each with 
$p$ D3 charge. What determines the most effective 
description depends on the region of parameter space.
For $N/g p^2 >>1$ the D5 brane description is the most 
favorable, and for $N/g p^2 <<1$ the NS5 brane takes over.

\subsection{\boldmath{$\th$}-angle Dependence}

We absorb the common phase of the fermion masses
$m$ and $m'$ into 
$\th = 2\pi{\rm Re}[\tau]$ such that we 
always take $m\in {\bf R}^+$.

Only five branes with NS charge are sensitive to the 
real part of $\tau$
\footnote{This is in accordance with the picture 
of the NS5 brane and world-sheet instantons being 
the dual descriptions of the confinement phase and instanton
effects, since these are $\th$-dependent
dynamical effects in field theory.}. For 
$(q,r)$ five branes, where $q=N/p$ and $r=0,...,p-1$,
we can write $M= p \tilde{\tau}$, 
with $\tilde{\tau}=(q\tau+r)/p$. 
As was observed in \cite{Aharony:2000nt},
the Polchinski-Strassler solution only depends 
on this effective modular parameter. In particular,
the five brane potential \eqref{5pot} becomes
\begin{equation}
\begin{split}
V_{NS5}(x,\vf) = \left(\frac{m^4 N^2}{16\pi}\right)
{\rm Im}\l[\frac{-1}{\tilde \tau}\r]x^2 
\left\{x^2 -2 x\,{\rm Re}\left[e^{-i\vf} 
+e^{3i\vf}\frac{b}{3}\left(\frac{\tilde{\tau}^\star}
{\tilde{\tau}}\right)^2\right]
\right.
\\
\left. +1+\frac{|b|^2}{3} 
-{\rm Re}\left[e^{2i\vf}c\frac{\tilde{\tau}^\star}
{\tilde{\tau}}\right]\right\} \,.
\end{split}
\label{NS5}
\end{equation}

As can be seen from the expressions for the potential
and the location of its critical points,
the precise analysis of the $\th$ dependence can be 
highly intricate. But we can make some general qualitative
observations. Since generically the value of the five brane potential
at the minimum is negative, the most favorable would be 
the $(q,r)$ five brane with the largest value of 
${\rm Im}\l[\frac{-1}{\tilde \tau}\r]$. 
As we already observed, this is the same as the criteria for the 
validity of the classical supergravity description in 
the $(q,r)$ five brane background, which to be concrete is
\begin{equation}
1<<{\rm Im}\l[\frac{-1}{\tilde \tau}\r]<<p^2 \,.
\end{equation}
Then, large $p$, and therefore ``small'' $q=N/p$, are preferred.
Furthermore, given $q$, the smallest vacuum energy would be 
for that $r$ for which $|\frac{\th}{2\pi} +r\frac{p}{N}|$
is smallest.

\subsection{Vacuum Structure for \boldmath{$\th=0$}}

Since $\th=0$, we select $r=0$ for the $(q,r)$ five brane
with $q\not=0$. From \eqref{5pot}, we see that 
for a $(q,0)$ or a $(p,0)$ five brane, the vacuum 
energy is suppressed by a $1/q^2$ or $1/p^2$ factor, respectively.
It is clear that $q=1$ (one NS5 brane; complete confinement)
or $p=1$ (one D5 brane; complete Higgsing) are selected.
Then, the vacuum structure is reduced to the analysis of
these two ($S$-dual) cases. 

But even then, the vacuum structure has 
a non-trivial dependence on the complex phases of 
the parameters $b$ and $c$. $R$-symmetry selection rules
give the $R$-charges: $R[z]=2/3$, $R[b]=-8/3$ and $R[c]=-2$.
This explains the combinations $e^{3i\vf}b$ and $e^{2i\vf}c$
appearing in \eqref{5pot} and \eqref{eq:PQ}.
If we want a five brane
wrapped on a two sphere with non-zero size 
we should require a positive discriminant for 
the real part of equation \eqref{eq:ext}:
\begin{equation}
\Delta = {\rm Re}[P]^2 -4{\rm Re}[Q] \geq 0 \,.
\label{discr}
\end{equation}
This dependence is complicated.
For fixed modulus $|b|$ and $|c|$,
there may be phases of $b$ and
$c$ which do not admit a positive discriminant.
In particular, the overall sign of the last term
in Q (see \eqref{eq:Q}) becomes crucial. 
At $\th=0$, this sign changes under $S$-duality.
To see the consequences of this duality, we can 
simplify our analysis by taking $b$ and $c$ to be real.

For real parameters $b$ and $c$, the phase $\vf$ vanishes.
The equation for the critical points 
simplifies to the real solutions of
\begin{equation}
x^2 -\frac{1}{2}(3+b) x +\frac{1}{2}\l(1+\frac{b^2}{3} 
-(-1)^{ns}c \r) =0 \,,
\label{eq:real_x}
\end{equation}
where $ns=0$ for the D5 brane and $ns=1$ for the NS5 brane.
Given a point in the space of
parameters $\{N,g,b,c\}$, the vacuum will be described by 
the five brane with the most negative vacuum energy,
\begin{equation}
V_{ns}(x) = \left(\frac{m^4 N^2}{16\pi}\right)
N g^{2ns-1}x^2 \left\{x^2 -2 x\,\l(1+\frac{b}{3}\r)
+1+\frac{|b|^2}{3} -(-1)^{ns}c\right\} \,.
\label{real5pot}
\end{equation}

\begin{figure}
 \centering  \mbox{\includegraphics[width=5in]{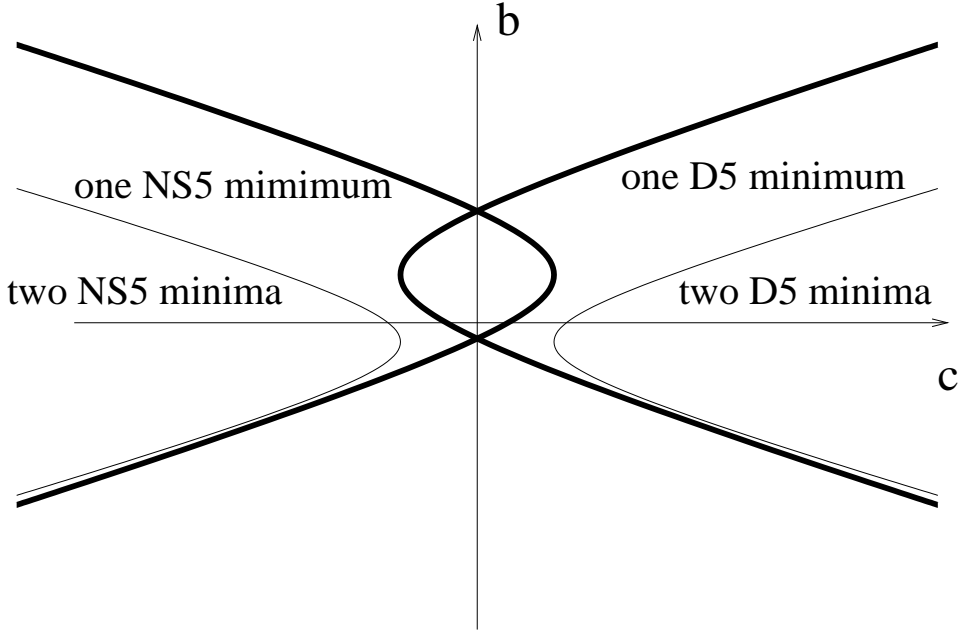}}
   \caption{Phase structure within PS approximation.}
   \label{fig:phase}
\end{figure}

Observe that the two potentials and their respective 
critical points are related 
by the interchange of $g \leftrightarrow 1/g$
and $c \leftrightarrow -c$. These correspond to 
mutually exclusive regions. The transformation 
involving the string 
coupling $g$ is clearly a consequence of $S$-duality
in $\cN=4$ SYM. The reason for the flip of sign in $c$
can be understood by the following: for real parameters,
the D5 brane world-volume corresponding to the two sphere
is oriented in the $\{x^7,x^8,x^9\}$ subspace, as can be seen
from \eqref{eq:z}; on the other hand, for the NS5 brane, 
it is spanned in the $\{x^4,x^5,x^6\}$ directions. 
Both orientations
are related by a $\pi/2$ rotation of \eqref{eq:z}, which 
is equivalent as flipping the sign of $c$.

When $c=0$, there is no distinction between the D5 and NS5
branes with respect to the location of the two sphere.
The only difference rests on the global $g$ 
dependence in the potential \eqref{real5pot}.
When $g<1$, the D5 vacuum is selected.
For $g>1$, the NS5 brane vacuum is the chosen one.
This is in complete agreement with the underlying 
$S$-duality of the unbroken supersymmetric theory.
A numerical analysis shows that only for 
$0\leq b \leq 3$ there is
a non-zero solution, $x_{min}\not=0$, of \eqref{eq:real_x} 
where the value of the five brane potential evaluated at $x_{min}$
is the absolute minimum.
\footnote{There is always a local minima
at $x=0$ with zero vacuum energy.}

When $c>0$, the region of the parameter $b$ with
a stable two sphere for the NS5 brane
monotonically reduces as $c$ is increased.
In fact, for $c>4/5$, there is no non-zero 
solution of \eqref{eq:real_x} for the NS5 brane;
it collapses to zero size, which lies outside 
of our approximation regime. 
On the other hand, for the D5 brane, 
increasing $c>0$ increases the range of the parameter $b$.
For $c\to \infty^+$, we have $|b_{max}| \to \frac{2}{5}\sqrt{30 c}$.

When $c<0$, the situation is completely reversed with respect to the 
D5 and NS5 branes. Negative $c$ 
defines the region favorable for the 
NS5 brane. 

\begin{figure}
\noindent
\begin{minipage}[b]{.46\linewidth}
\centering\includegraphics[width=\linewidth]{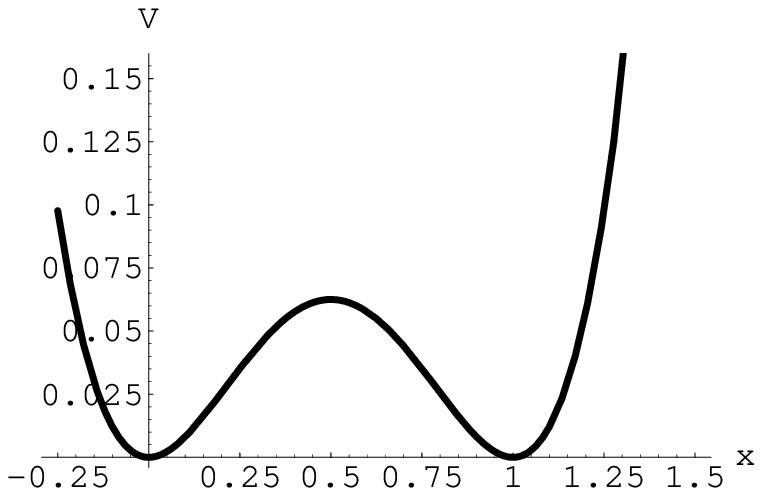}
\caption{Supersymmetric potential.}
\label{fig:susy}
\end{minipage}\hfill
\begin{minipage}[b]{.46\linewidth}
\centering\includegraphics[width=\linewidth]{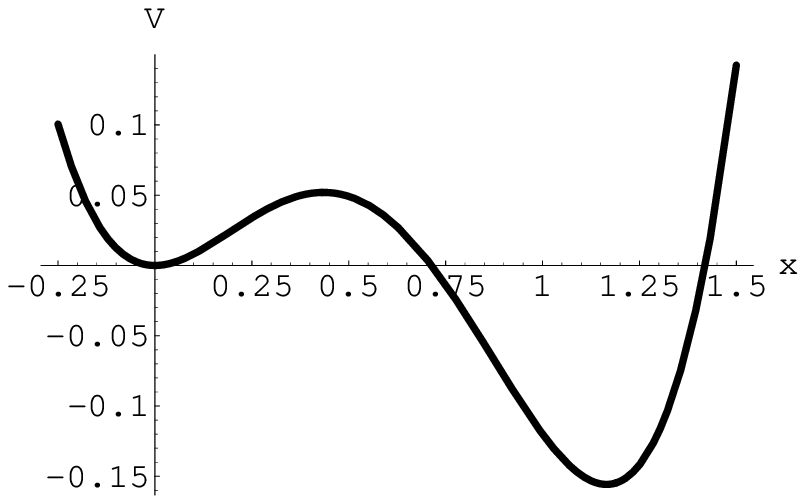}
\caption{Phase with only one five brane minimum.}
\label{fig:max}
\end{minipage}
\end{figure}

\begin{figure}
\begin{minipage}[b]{.46\linewidth}
\centering\includegraphics[width=\linewidth]{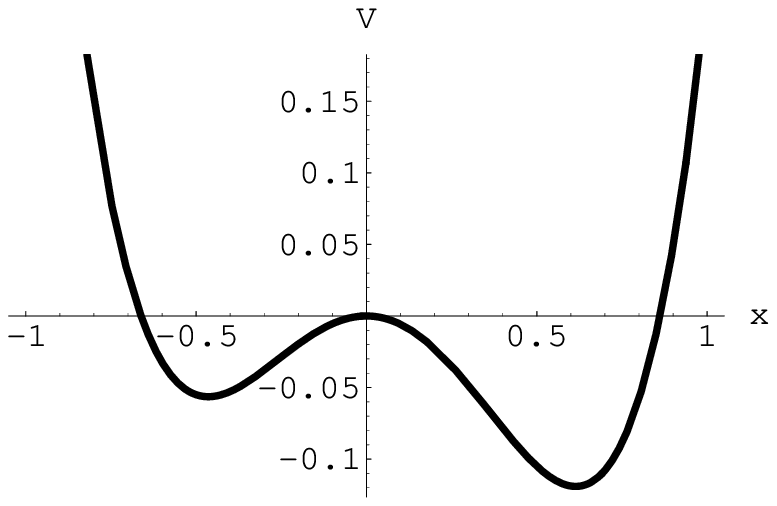}
\caption{Phase with two five brane minima.}
\label{fig:min}
\end{minipage}\hfill
\begin{minipage}[b]{.46\linewidth}
\centering\includegraphics[width=\linewidth]{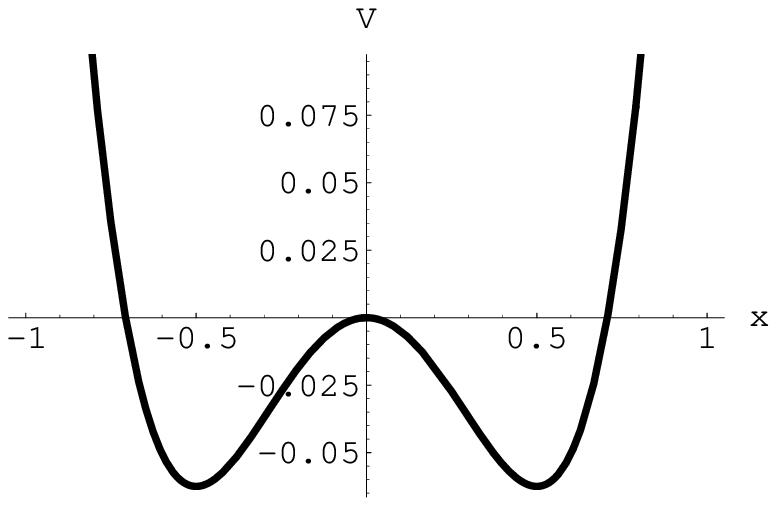}
\caption{Potential at $b=-3$.}
\label{fig:b3}
\end{minipage}
\end{figure}

The global picture is presented in figure 1. 
When $b=c=0$, which is the $\cN =1^*$ supersymmetric point,
the effective potential \eqref{real5pot} looks like 
that of figure \ref{fig:susy}; the five brane sits at the 
minimum with non-zero $x$ and the vacuum energy is zero.
Close to the $\cN =1^*$ point, 
when $b$ and $c$ are smaller than one, 
what determines the vacuum is essentially the 
value of the string coupling, $g$.
For $g<1$ ($g>1$) the D5 (NS5)
brane vacuum is selected.
Far from the origin, two symmetric 
branches appear: the Higgs branch for $c>0$ and the confining
branch for $c<0$. In each of the branches, 
there is a critical line,
defined by the vanishing of the smaller 
root of \eqref{eq:real_x}, $x_-(b,c)=0$.
This line divides the branch into two regions. 
In one region, $x_-$ is a maximum and the potential
typically looks like in figure \ref{fig:max}.
In the other region, $x_-$ is a local minimum,
and the potential becomes like in figure \ref{fig:min}.
The minimum located at the largest value of the AdS radius, 
given by the bigger root $x_+$, is the absolute one. 
The exception is when $b=-3$, 
where both minima collide at the same absolute value of $x$
and we just have a potential like that of figure \ref{fig:b3}.

Notice that we can scale the $\cN=1$ preserving masses to zero
and keep non-zero $m'$ and $\mu^2$ supersymmetry breaking 
parameters. It corresponds to the asymptotic regions
of the phase space shown in figure 1.
Take the limit $\epsilon \to 0$,
keeping fixed
\begin{equation}
{\tilde m}= \frac{m}{\epsilon}\ ,\quad
{\tilde b} = b\epsilon \ ,\quad {\tilde c}=c\epsilon^2 
\ ,\quad {\tilde x}=x\epsilon \,.
\end{equation}
The five brane location \eqref{eq:z} is kept fixed and
different from zero 
as far as $\tilde{b}^2\leq (-1)^{ns} \frac{24}{5}\tilde{c}$.
Finally, the effective potential \eqref{real5pot} becomes
\begin{equation}
V_{ns}({\tilde x})= \left(\frac{{\tilde m}N^2}{16\pi}\right)
N g^{2ns-1}{\tilde x}^2 \left\{
{\tilde x}^2 
-\frac{2{\tilde b}}{3}{\tilde x}+
\frac{{\tilde b}^2}{3} -(-1)^{ns}{\tilde c}\right\} \,.
\end{equation}

\bigskip
%%%%%%%%%%%%%%%%%%%%%%%%%%%%%%%%%%%%%%%%%%%%%%%%%%%%%%%%%%%%%

\section{Discussion}

In this paper we have analyzed the $SO(3)$-invariant
region of parameter space  
where the PS approximation is still valid.
By this we mean the region where the relevant 
effective potential is given by expression
\eqref{5pot_2}, with its minimum given
by stable five brane/D3 states 
wrapped on a round two-sphere of finite size.
Beyond this region, either the potential 
\eqref{5pot_2} ceases to be a good approximation 
or the five brane collapses. Therefore, any significant
extension of its limits would require qualitative 
(and probably quantitative) new work.
In particular, it would be very interesting to reach the 
$SU(3)$ invariant point $m=\mu^2=0$, $m'\not=0$
and to find out if the five-dimensional
singularity found in \cite{Distler:1999tr} 
\footnote{which satisfies the criteria established in 
\cite{Gubser:2000nd}.} is resolved 
and we have chiral symmetry breaking to $SO(3)$. 

Let us add some comments on the field theory physics.
Most of the qualitative features exposed in 
\cite{Polchinski:2000uf} continue to hold
here. For instance, the $(p,q)$-flux tubes were realized as 
$(p,q)$ one-five/D3 near BPS bound states. 
Since our parameters $m'$ and $\mu^2$ break
supersymmetry softly, we believe 
that these bound states are still there, 
at least in the PS approximation region. 
To compute their IR string tension, we need the 
near-shell background for the five brane metric and 
the dilaton. Qualitatively, they are given by expressions 
similar to those in \cite{Polchinski:2000uf}, 
with the main difference 
being the location of the five brane along the 
AdS radial direction.
The same can be said about the field theory spectrum,
for instance the construction of baryons.
Since only one vacuum is selected, the domain walls are 
lost. With respect to the condensates, there is a new 
dependence on the $m'$ and $\mu^2$ parameters.
In particular, there is a new non-zero condensate 
related to the operator ${\rm tr}(F^2)$
\footnote{We remind that 
without significant supersymmetry constraints,
there is mixing between composite operators.
See \cite{Aharony:2000nt} for a discussion in the 
context of the $\cN=1^\star$ theory.}, which can 
be obtained from the normalizable mode corresponding
to the $SU(3)$ singlet.

And finally, some words of caution are in order
regarding the starting UV fixed point. 
If we want to restrict
the analysis to classical supergravity, we have to take
(at least) large N and strong bare 't Hooft coupling. 
Then, all the fields in the $\cN=4$ supermultiplet
continue to be relevant at the RG-invariant generated scale.
As noticed in \cite{Dorey:2000fc},
there are qualitative new features in the $\cN=1^\star$
vacua with respect the $\cN=1$ SYM vacua.
Decoupling of the additional $\cN=4$ fields
involves decreasing the 't Hooft coupling,
whose inverse parameter controls the $\alpha'$ corrections
in the AdS/CFT correspondence.

\bigskip
%%%%%%%%%%%%%%%%%%%%%%%%%%%%%%%%%%%%%%%%%%%%%%%%%%%%%%%%%%%%
\section*{Acknowledgements}

I thank J. Distler and M. Strassler for illuminating
discussions and E. Gorbatov for a critical reading 
of the manuscript.
This work has been supported by the NSF Grant PHY9511632 and
the Robert A.~Welch Foundation.

\bigskip
%%%%%%%%%%%%%%%%%%%%%%%%%%%%%%%%%%%%%%%%%%%%%%%%%%%%%%%%%%%%%%

\renewcommand{\baselinestretch}{.7} \normalsize
\bibliography{ps}
\bibliographystyle{utphys}

%%%%%%%%%%%%%%%%%%%%%%%%%%%%%%%%%%%%%%%%%%%%%%%%%%%%%%%%%%%%%

\end{document}